\documentclass[12pt]{article}
\usepackage{graphics}
\usepackage{amssymb}

\newcommand{\QED}{\mbox{\rule[-1.5pt]{6pt}{10pt}}}
\newtheorem{claim}{Claim}[section]
\newtheorem{theorem}[claim]{Theorem}
\newtheorem{proposition}[claim]{Proposition}
\newtheorem{lemma}[claim]{Lemma}
\newtheorem{corollary}[claim]{Corollary}

\begin{document}

\title{Band gap of the Schr{\"o}dinger operator
 with a strong $\delta$-interaction on a periodic curve}
\author{P.~Exner$^{a,b}$ and K.~Yoshitomi$^{c}$}
\date{}
\maketitle
\begin{quote}
{\small \em a) Department of Theoretical Physics, Nuclear Physics
Institute, \\ \phantom{e)x}Academy of Sciences, 25068 \v Re\v z,
Czech Republic \\
 b) Doppler Institute, Czech Technical University,
 B\v{r}ehov{\'a} 7,\\
\phantom{e)x}11519 Prague, Czech Republic \\
 c) Graduate School of Mathematics, Kyushu University,
Hakozaki, \\ \phantom{e)x}Fukuoka 812-8581, Japan \\
 \rm \phantom{e)x}exner@ujf.cas.cz,
 yositomi@math.kyushu-u.ac.jp}
\vspace{8mm}

\noindent {\small In this paper we study the operator
$H_{\beta}=-\Delta-\beta\delta(\cdot-\Gamma)$ in
$L^{2}(\mathbb{R}^{2})$, where $\Gamma$ is a smooth periodic curve
in $\mathbb{R}^{2}$. We obtain the asymptotic form of the band
spectrum of $H_{\beta}$ as $\beta$ tends to infinity. Furthermore,
we prove the existence of the band gap of $\sigma(H_{\beta})$ for
sufficiently large $\beta>0$. Finally, we also derive the spectral
behaviour for $\beta\to\infty$ in the case when $\Gamma$ is
non-periodic and asymptotically straight.}
\end{quote}


\section{Introduction}

In this paper we are going to discuss some geometrically induced
spectral properties of singular Schr\"odinger operators which can
be formally written as $H_{\beta}= -\Delta
-\beta\delta(\cdot-\Gamma)$, where $\Gamma$ is an infinite curve
in the plane.

This problem stems from physical interest to quantum mechanics of
electrons confined to narrow tubelike regions usually dubbed
``quantum wires''. Such systems are often modeled  by means of
Schr\"odinger operators on curves, or more generally, on graphs.
This is an idealization, however, because in reality the electrons
are confined in a potential well of a finite depth, and therefore
one can find them also in the exterior of such a ``wire'', even if
not too far since this a classically forbidden region. The
generalized Schr\"odinger operators mentioned above provide us
with a simple model which can take such tunneling effects into
account.

Singular interactions have been studied by numerous authors -- see
the classical monograph \cite{AGHH}, and the recent volume
\cite{AK} for an up-to-date bibliography. While the general
concepts are well known, the particular case of a
$\delta$-interaction supported by a curve attracted much less
attention; we can mention \cite{BT, BEKS} and a recent article
\cite{EI}, where a nontrivial relation between spectral properties
and the geometry of the curve $\Gamma$ was found for the first
time. It was followed by our previous paper \cite{EY}, where we
posed the question about the strong coupling asymptotic behaviour,
$\beta\to \infty$, of the eigenvalues of $H_{\beta}$ in the case
when $\Gamma$ was a loop. We have shown there that the asymptotics
is given by the spectrum of the Schr\"odinger operator on
$L^2(\Gamma)$ with a curvature-induced potential.

Here we are going to discuss a similar problem in the situation
when $\Gamma$ is an infinite smooth curve without
self-intersections. We pay most attention to the case of a
periodic $\Gamma$ where we find the asymptotic form of the
spectral bands and prove existence of open band gaps for $\beta>0$
large enough provided $\Gamma$ is not a straight line. We also
treat the case of a non-straight $\Gamma$ which is straight
asymptotically, and thus by \cite{EI} it gives rise to a nonempty
discrete spectrum; we find the behaviour of these eigenvalues for
$\beta\to \infty$.

While the basic idea is the same as in \cite{EY}, namely
combination of a bracketing argument with the use of suitable
curvilinear coordinates in the vicinity of $\Gamma$, the periodic
case requires several more tools. Let us review briefly the
contents of the paper. In the following section we present a
formulation of the problem and state the results. Section~3 is
devoted to the proof of our main result, Theorem~\ref{main}. We
perform the Floquet-Bloch reduction and estimate the discrete
spectrum of the fiber operator $H_{\beta,\theta}$ using a
Dirichlet-Neumann bracketing and approximate operators with
separated variables. As a corollary we obtain the existence of
open gaps for $\beta$ large enough. To get a more specific
information on the last question, we derive in Section~4 a
sufficient condition under which the $n$th gap is open for a given
$n$. The final section deals with the case of an asymptotically
straight
 $\Gamma$.


\setcounter{equation}{0}
\section{Main results}

Let us first introduce the needed notation and formulate the
problem. The main topic of this paper is the Schr{\"o}dinger
operator with a $\delta$-interaction on a periodic curve. Let
$\Gamma:\mathbb{R}\owns s\mapsto (\Gamma_{1}(s),\Gamma_{2}(s))\in
\mathbb{R}^{2}_{x,y}$ be a curve which is parametrized by its arc
length. Let $\gamma:\mathbb{R}\rightarrow\mathbb{R}$ be the signed
curvature of $\Gamma$, i.e. $\gamma(s):=(\Gamma^{\prime\prime}_{1}
\Gamma^{\prime}_{2}-\Gamma^{\prime\prime}_{2}\Gamma^{\prime}_{1})(s)$.
We impose on it the following assumptions:
 \begin{description}
\vspace{-.6ex}
\item{(A.1)} $\gamma\in C^{2}(\mathbb{R})$.
\vspace{-1.2ex}
\item{(A.2)} There exists $L>0$ such that
 $\gamma(\cdot+L)=\gamma(\cdot)$ on $\mathbb{R}$.
\vspace{-1.2ex}
\item{(A.3)} $\displaystyle{\int^{L}_{0}}\gamma(t)\,dt=0.$
\vspace{-.6ex}
\end{description}
Given $\beta>0$, we define
 $$ q_{\beta}(f,f)=\Vert\nabla
f\Vert^{2}_{L^{2}(\mathbb{R}^{2})}
-\beta\int_{\Gamma}|f(x)|^{2}\,dS \quad\mathrm{for}\quad f\in
H^{1}(\mathbb{R}^{2}). $$
By $H_{\beta}$ we denote the self-adjoint operator associated with
the form $q_{\beta}$. The operator $H_{\beta}$ can be formally
written as $-\Delta-\beta \delta (\cdot-\Gamma)$. Our main purpose
is to study the asymptotic behaviour of the band spectrum of
$H_{\beta}$ as $\beta$ tends to infinity. Let $\alpha\in [0,2\pi)$
be the angle between the vectors $\Gamma^{\prime}(0)$ and $(1,0)$:
$\Gamma^{\prime}(0)=(\cos\alpha,\sin\alpha)$. We define new
coordinates $(x^{\prime},y^{\prime})$ by
 $$\left(
 \matrix{
 x^{\prime}\cr
 y^{\prime}\cr}
 \right)=
 \left(
 \matrix{\cos\alpha&\sin\alpha\cr
 -\sin\alpha&\cos\alpha\cr}
 \right)
 \left(
 \matrix{
 x-\Gamma_{1}(0)\cr
 y-\Gamma_{2}(0)\cr}
 \right).$$
From now on, we work in the coordinates $(x^{\prime},y^{\prime})$,
where the curve $\Gamma$ assumes the form
 \begin{eqnarray*}
\Gamma_{1}(s) &=& \int^{s}_{0}\cos
\left(-\int^{t}_{0}\gamma(u)\,du \right)\,dt, \\ \Gamma_{2}(s) &=&
\int^{s}_{0}\sin \left(-\int^{t}_{0}\gamma(u)\,du\right)\,dt.
 \end{eqnarray*}
Combining these relations with $(A.3)$, we have
 \begin{equation} \label{1.1}
\Gamma(\cdot+L)-\Gamma(\cdot)=(K_{1},K_{2})\quad
\mathrm{on}\quad\mathbb{R},
\end{equation}
where
 \begin{eqnarray*}
K_{1} &=& \int^{L}_{0}\cos \left(-\int^{t}_{0}\gamma(u)\,du
\right)\,dt, \\ K_{2} &=& \int^{L}_{0}\sin
\left(-\int^{t}_{0}\gamma(u)\,du \right)\,dt.
 \end{eqnarray*}
In the vicinity of $\Gamma$ one can introduce the natural locally
orthogonal system of curvilinear coordinates. By $\Phi$ we denote
the map
 $${\mathbb R}^{2}\owns (s,u)\mapsto
 (\Phi_{1}(s,u),\Phi_{2}(s,u))=(\Gamma_{1}(s)-u\Gamma^{\prime}_{2}(s),
\Gamma_{2}(s)+u\Gamma^{\prime}_{1}(s))\in{\mathbb R}^{2}.$$
We further impose the following assumptions on $\Gamma$:
 \begin{description}
 \vspace{-.6ex}
\item{(A.4)} $K_{1}>0.$
\vspace{-1.2ex}
\item{(A.5)} There exists $a_{0}>0$ such that the map
 $\Phi|_{[0,L)\times (-a,a)}$ is injective and $\Phi((0,L)\times
(-a,a))\subset(0,K_{1})\times {\mathbb R}$ for all $a\in
(0,a_{0})$.
\end{description}
As in the proof of [Yo, Proposition 3.5], we notice that the
assumptions $(A.4)$ and $(A.5)$ are satisfied, e.g., if
$\max_{t\in [0,L]}|\int^{t}_{0}\gamma (s)\,ds|<\pi/2$; on the
other hand, this condition is by no means necessary. Let us also
remark that in general the choice of the initial point $s=0$ is
important in checking the assumptions $(A.4)$ and $(A.5)$. We put
 $$\Lambda=(0, K_{1})\times {\mathbb R}.$$
For $\theta\in[0,2\pi)$, we define
 \begin{eqnarray*}
Q_{\theta} &=& \{u\in H^{1}(\Lambda);\quad u(K_{1},
K_{2}+\cdot)=e^{i\theta}u(0,\cdot)\quad\mathrm{on}\quad\mathbb{R}\},
\\ q_{\beta,\theta}(f,f) &=& \Vert\nabla
 f\Vert^{2}_{L^{2}(\Lambda)}
-\beta\int_{\Gamma((0,L))}|f(x)|^{2}\,dS \quad\mathrm{for}\quad
f\in Q_{\theta}.
 \end{eqnarray*}
By $H_{\beta,\theta}$ we denote the self-adjoint operator
associated with the form $q_{\beta,\theta}$. We shall prove in
Lemma~\ref{lemma1} the unitary equivalence
 \begin{equation}
H_{\beta}\cong\int^{2\pi}_{0}\oplus H_{\beta,\theta}\,d\theta.
\label{1.2} \end{equation}
By Lemma~\ref{lemma3} this implies
 \begin{equation}
\sigma(H_{\beta})= \bigcup_{\theta\in[0,2\pi)}
\sigma(H_{\beta,\theta}).\label{1.3}
\end{equation}
Since $\Gamma((0,L))$ is compact, we infer by Lemma~\ref{lemma2}
that
 \begin{equation}
\sigma_{\mathrm{ess}}(H_{\beta,\theta})=[0,\infty).\label{1.4}
\end{equation}
Next we need a comparison operator on the curve. For a fixed
$\theta\in [0,2\pi)$ we define
 $$ S_{\theta}=-\frac{d^{2}}{d s^{2}}-\frac{1}{4}\gamma(s)^{2}
\quad\mathrm{in}\quad L^{2}((0,L))$$
with the domain
 $$ P_{\theta}=\{ u\in H^{2}((0,L));\quad
u(L)=e^{i\theta}u(0),\quad
u^{\prime}(L)=e^{i\theta}u^{\prime}(0)\}.$$
For $j\in\mathbb{N}$, we denote by $\mu_{j}(\theta)$ the $j$th
eigenvalue of the operator $S_{\theta}$ counted with multiplicity.
This allows us to formulate our main result.

 \begin{theorem} \label{main}
Let $n$ be an arbitrary integer. There exists $\beta(n)>0$ such
that
$$ \sharp\sigma_\mathrm{d}(H_{\beta,\theta})\geq n
\quad\mathrm{for}\quad \beta\geq \beta(n) \quad\mathrm{and}\quad
\theta\in [0,2\pi).$$
For $\beta\geq \beta(n)$ we denote by $\lambda_{n}(\beta,\theta)$
the $n$th eigenvalue of $H_{\beta,\theta}$ counted with
multiplicity. Then $\lambda_{n}(\beta,\theta)$ admits an
asymptotic expansion of the form
$$ \lambda_{n}(\beta,\theta)=-\frac{1}{4}\beta^{2}
+\mu_{n}(\theta)+\mathcal{O}(\beta^{-1}\log\beta)
\quad\mathrm{as}\quad\beta \rightarrow\infty,\label{1.5} $$
where the error term is uniform with respect to $\theta\in
[0,2\pi)$.
 \end{theorem}
Combining this result with Borg's theorem on the inverse problem
for Hill's equation, we obtain the following corollary about the
existence of the band gap of $\sigma (H_{\beta})$.
 \begin{corollary} \label{gap}
Assume that $\gamma\neq 0$, i.e. that $\Gamma$ is not a straight
line. Then there exists $m\in\mathbb{N}$ and $G_{m}>0$ such that
$$ \lim_{\beta\rightarrow\infty}\left( \min_{\theta\in
[0,2\pi)}\lambda_{m+1}(\beta,\theta) -\max_{\theta\in
[0,2\pi)}\lambda_{m}(\beta,\theta)\right) =G_{m}. $$
 \end{corollary}
\vspace{1em}

\noindent We would like to know, of course, which gaps in the
spectrum open as $\beta\to\infty$. To this aim we prove a
sufficient condition which guarantees this property for a fixed
gap index $n$. Let $\{c_{j}\}^{\infty}_{j=1}$ and
$\{d_{j}\}^{\infty}_{j=0}$ be the Fourier coefficients of
$\frac{1}{4}\gamma (s)^{2}$:
 \begin{equation}
 \frac{1}{4}\gamma (s)^{2}=\sum^{\infty}_{j=1}
c_{j}\sin\frac{2\pi j}{L}s+\sum^{\infty}_{j=0}d_{j} \cos\frac{2\pi
j}{L}s\quad\mathrm{in}\quad L^{2}((0,L)). \label{fourier}
 \end{equation}
 \begin{proposition} \label{gap-cond}
Let $n\in\mathbb{N}$. Assume that $\,0<\sqrt{c^{2}_{n}+d^{2}_{n}}
<\frac{12\pi^{2}}{L^{2}}n^{2}$ and
$$ \max_{s\in [0,L]}\left| \frac{1}{4}\gamma(s)^{2}
-d_{0}-c_{n}\sin\frac{2n\pi}{L}s-d_{n}\cos\frac{2n\pi}{L}s\right|
<\frac{1}{4}\sqrt{c^{2}_{n}+d^{2}_{n}}, $$
then we have
$$ \lim_{\beta\rightarrow\infty}\left( \min_{\theta\in
[0,2\pi)}\lambda_{n+1}(\beta,\theta) -\max_{\theta\in
[0,2\pi)}\lambda_{n}(\beta,\theta)\right)
>0. $$
 \end{proposition}
\vspace{1em}

\noindent In particular, it is obvious that if the effective
curvature-induced potential has a dominating Fourier component in
the expansion (\ref{fourier}), the band with the same index opens
as $\beta\to \infty$. We also see that the second assumption of
Proposition~\ref{gap-cond} is more difficult to satisfy as the
index $n$ increases.


\setcounter{equation}{0}
\section{Proof of Theorem~\ref{main}}

We first prove the unitary equivalence $(2.2)$ by using the
Floquet-Bloch reduction scheme -- see, e.g., [RS, XIII.16]. For
$u\in C^{\infty}_{0}(\mathbb{R}^{2})$ and $\theta\in [0,2\pi)$, we
define
$$ \mathcal{U}_{0}u(x,y,\theta)=
\frac{1}{\sqrt{2\pi}}\sum^{\infty}_{m=-\infty}
e^{im\theta}u(x-mK_{1},y-mK_{2}),\quad (x,y)\in\Lambda. $$
Then $\mathcal{U}_{0}$ extends uniquely a unitary operator from
$L^{2}(\mathbb{R}^{2})$ to $\int^{2\pi}_{0} \oplus
L^{2}(\Lambda)\,d\theta$, which we denote as $\mathcal{U}$. In
addition, $\mathcal{U}$ is unitary also as an operator from
$H^{1}(\mathbb{R} ^{2})$ to $\int^{2\pi}_{0}\oplus
H^{1}(\Lambda)d\theta$. Let us check the following claim.
 \begin{lemma} \label{lemma1}
We have
 \begin{equation}
 \mathcal{U}H_{\beta}\,\mathcal{U}^{-1}=\int^{2\pi}_{0}\oplus
H_{\beta,\theta}\,d\theta.\label{3.1} \end{equation}
 \end{lemma}
{\sl Proof:} We shall first show that
 \begin{equation}
q_{\beta}(f,g)=\int^{2\pi}_{0}q_{\beta,\theta}((
\mathcal{U}f)(\cdot,\cdot,\theta),
(\mathcal{U}g)(\cdot,\cdot,\theta)
)\,d\theta\quad\mathrm{for}\quad f,g\in
H^{1}(\mathbb{R}^{2}).\label{3.2} \end{equation}
Let $u,v\in C^{\infty}_{0}(\mathbb{R}^{2})$. The quadratic form
$$ q_{\beta}(u,v)= (\nabla u,\nabla v)_{L^{2}(\mathbb{R}^{2})}
-\beta\int_{\Gamma}u(x)\overline{v(x)}\,dS $$
can be in view of $(2.1)$ written as
 \begin{eqnarray*}
\lefteqn{ \sum^{\infty}_{m=-\infty}
((\nabla u)(x-mK_{1},x-mK_{2}),
(\nabla v)(x-mK_{1},y-mK_{2}))_{L^{2}(\Lambda)}} \\ &&
-\beta\sum^{\infty}_{m=-\infty}\int_{\Gamma((0,L))}
u(x-mK_{1},y-mK_{2})\overline{v(x-mK_{1},y-mK_{2})}\,dS
 \end{eqnarray*}
and since $\{\frac{1}{\sqrt{2\pi}}e^{in\theta}\}^{\infty}
_{n=-\infty}$ is a complete orthonormal system of
 $L^{2}((0,2\pi))$ we have
  \begin{eqnarray}
&=& \int^{2\pi}_{0}\left(\frac{1}{\sqrt{2\pi}}
\sum^{\infty}_{m=-\infty} e^{im\theta}(\nabla
u)(x-mK_{1},y-mK_{2}),\right. \nonumber \\&&
\left.\frac{1}{\sqrt{2\pi}}\sum^{\infty}_{n=-\infty}
e^{in\theta}(\nabla v)(x-nK_{1},y-nK_{2})
\right)_{L^{2}(\Lambda)}\, d\theta \nonumber \\&& -\beta
\int^{2\pi}_{0}\left(\frac{1}{\sqrt{2\pi}}
\sum^{\infty}_{m=-\infty} e^{im\theta}
u(x-mK_{1},y-mK_{2}),\right. \nonumber \\&&
\left.\frac{1}{\sqrt{2\pi}}\sum^{\infty}_{n=-\infty}
e^{in\theta}v(x-nK_{1},y-nK_{2}) \right)_{L^{2}(\Gamma
((0,L)))}\, d\theta \nonumber \\ &=&
\int^{2\pi}_{0}q_{\beta,\theta}((\mathcal{U}u)
(\cdot,\cdot,\theta),
(\mathcal{U}v)(\cdot,\cdot,\theta))\,d\theta.\label{3.3}
 \end{eqnarray}
Let $f,g\in H^{1}(\mathbb{R}^{2})$. Since
$C^{\infty}_{0}(\mathbb{R} ^{2})$ is dense in
$H^{1}(\mathbb{R}^{2})$, we can choose in it two sequences
$\{u_{j}\}^{\infty}_{j=1}$ and $\{v_{j}\}^{\infty}_{j=1}$ such
that
$$ u_{j}\rightarrow f\quad\mathrm{in}
\quad H^{1}(\mathbb{R}^{2}),
\quad v_{j}\rightarrow g\quad\mathrm{in}\quad
H^{1}(\mathbb{R}^{2}) \quad\mathrm{as}\quad
 j\rightarrow\infty.$$
The form $q_{\beta}$ is bounded in
 $H^{1}(\mathbb{R}^{2})$, hence
we get
 \begin{equation}
\lim_{j\rightarrow\infty}q_{\beta}(u_{j},v_{j})
=q_{\beta}(f,g).\label{3.4} \end{equation}
Notice that there exist a constant $C>0$ such that for any
$\theta\in [0,2\pi)$ and $u,v\in Q_{\theta}$, we have
 \begin{equation}
|q_{\beta,\theta}(u,v)|\leq C\Vert u
\Vert_{H^{1}(\Lambda)} \Vert
v\Vert_{H^{1}(\Lambda)}.\label{3.5} \end{equation}
Since $\mathcal{U}$ is a unitary operator from
$H^{1}(\mathbb{R}^{2})$ to $\int^{2\pi}_{0}\oplus
H^{1}(\Lambda)\,d\theta$, we have
 \begin{eqnarray*} \mathcal{U}u_{j}\rightarrow
\mathcal{U}f\quad &\mathrm{in}& \quad \int^{2\pi}_{0}\oplus
H^{1}(\Lambda)\,d\theta, \\ \mathcal{U}v_{j}\rightarrow
\mathcal{U}g\quad &\mathrm{in}& \quad \int^{2\pi}_{0}\oplus
H^{1}(\Lambda)\,d\theta.
 \end{eqnarray*}
Combining these relations with $(3.5)$, we have
 \begin{eqnarray}
 \lefteqn{\lim_{j\rightarrow\infty}\int^{2\pi}_{0}
q_{\beta,\theta}((\mathcal{U}u_{j})(\cdot,\cdot,\theta),
(\mathcal{U}v_{j})(\cdot,\cdot,\theta))
\,d\theta} \nonumber \\&&
=\int^{2\pi}_{0}q_{\beta,\theta}((\mathcal{U}f)
(\cdot,\cdot,\theta),
(\mathcal{U}g)(\cdot,\cdot,\theta))\,d\theta.\label{3.6}
 \end{eqnarray}
Putting (\ref{3.3}), (\ref{3.4}),
 and (\ref{3.6}) together, we get
(\ref{3.2}).

Next we shall show that
 \begin{equation}
\mathcal{U}^{-1} \left(\int^{2\pi}_{0}\oplus
H_{\beta,\theta}\,d\theta \right)\mathcal{U}\subset
H_{\beta}.\label{3.7}
\end{equation}
Let $u\in L^{2}(\mathbb{R}^{2})$ and $\mathcal{U}u\in
\mathcal{D}(\int^{2\pi}_{0}\oplus H_{\beta,\theta}\,d\theta)$. By
definition of the direct integral we have
 \begin{eqnarray}
&& (\mathcal{U}u)(\cdot,\cdot,\theta)\in
\mathcal{D}(H_{\beta,\theta})\quad\mathrm{for\quad a.e.}\quad
\theta\in [0,2\pi), \nonumber \\ && \int^{2\pi}_{0}\Vert
H_{\beta,\theta} \mathcal{U}u(\cdot,\cdot,\theta)
\Vert^{2}_{L^{2}(\Lambda)}\,d\theta<\infty.\label{3.8}
 \end{eqnarray}
The first named property means in particular that $(\mathcal{U}u)
(\cdot,\cdot,\theta)\in \mathcal{D}(H_{\beta,\theta})$ for a.e.
$\theta\in [0,2\pi)$, thus we have
 \begin{equation}
q_{\beta,\theta}((\mathcal{U}u)(\cdot,\cdot,\theta),g)
=(H_{\beta,\theta}\mathcal{U}u(\cdot,\cdot,\theta),g)
_{L^{2}(\Lambda)}
\quad\mathrm{for\quad all}\quad g\in Q_{\theta}.\label{3.9}
 \end{equation}
Note that there exists a constant $b>0$ such that for all
$\theta\in [0,2\pi)$ and $f\in Q_{\theta}$, we have
 \begin{equation}
q_{\beta,\theta}(f,f)+b\Vert f\Vert^{2}_{L^{2}(\Lambda)}
\geq\frac{1}{2}\Vert f\Vert^{2}_{H^{1}(\Lambda)}.\label{3.10}
\end{equation}
It follows from $(3.9)$ that
 \begin{eqnarray*}
\lefteqn{|q_{\beta,\theta}(\mathcal{U}u(\cdot,\cdot,\theta),
\mathcal{U}u(\cdot,\cdot,\theta))| = |(H_{\beta,\theta}\mathcal{U}
u(\cdot,\cdot,\theta), \mathcal{U}u(\cdot,\cdot,\theta))|} \\&&
\leq \frac{1}{2}\left(\Vert H_{\beta,\theta} \mathcal{U}
u(\cdot,\cdot,\theta)\Vert^{2}_{L^{2}(\Lambda)} +\Vert\mathcal{U}
u(\cdot,\cdot,\theta)\Vert^{2}_{L^{2}(\Lambda)}\right).
\end{eqnarray*}
This together with (\ref{3.8}) and (\ref{3.10}) implies that
$\mathcal{U}u \in\int^{2\pi}_{0}\oplus H^{1}(\Lambda)\,d\theta$,
so we have $u\in H^{1}(\mathbb{R}^{2})$. We pick any $v\in
H^{1}(\mathbb{R}^{2})$. Its image by $\mathcal{U}$ satisfies
$$ (\mathcal{U}v)(\cdot,\cdot,\theta)\in Q_{\theta}
\quad\mathrm{for\quad a.e.}\quad\theta\in [0,2\pi).$$
We put $w(\theta) =H_{\beta,\theta}
\mathcal{U}u(\cdot,\cdot,\theta)$. From $(3.2)$ we have
$$ q_{\beta}(u,v)= \int^{2\pi}_{0}
q_{\beta,\theta}((\mathcal{U}u)(\cdot,\cdot,\theta),
(\mathcal{U}v)(\cdot,\cdot,\theta))\,d\theta $$
which can be using (\ref{3.9}) rewritten as
$$ q_{\beta}(u,v)=
\int^{2\pi}_{0}(w(\theta),(\mathcal{U}v)(\cdot,\cdot,\theta))
_{L^{2}(\Lambda)}\,d\theta\\
=(\mathcal{U}^{-1}w,v)_{L^{2}(\mathbb{R}^{2})}. $$
Using $(3.8)$, we get
$$ \mathcal{U}^{-1}w\in L^{2}(\mathbb{R}^{2}).$$
Thus we have $u\in \mathcal{D}(H_{\beta})$ and
$$ \mathcal{U}^{-1}\left(\int^{2\pi}_{0}\oplus
H_{\beta,\theta}\,d\theta\right)\mathcal{U}u=H_{\beta}u,$$
which proves (\ref{3.7}). Since the two operators in this
 inclusion are self-adjoint, we arrive at (\ref{3.1}). \quad \QED
\vspace{1em}

Next we have to localize the essential spectrum of our operator.
 \begin{lemma} \label{lemma2}
We have
$$ \sigma_{\mathrm{ess}}(H_{\beta,\theta})=[0,\infty).$$
\end{lemma}
{\sl Proof:} We define
$$ c_{\theta}(u,v)=\int_{\Gamma((0,L))}u(x)\overline{v(x)}\,dS,
\quad u,v\in Q_{\theta},$$
which allows us to write $q_{\beta,\theta}=q_{0,\theta}-\beta
c_{\theta}$ on $Q_{\theta}$. Let $C_{\theta}$ be the self-adjoint
operator associated with the form $\overline{c_{\theta}}$. In view
of the quadratic form version of Weyl's theorem (see [RS, XIII.4,
Corollary 4]), it suffices to demonstrate that the operator
$(H_{0,\theta}+1)^{-1}C_{\theta}(H_{0,\theta}+1)^{-1}$ is compact
on $L^{2}(\Lambda)$. Let $\{u_{n}\}^{\infty}_{n=1}\subset
L^{2}(\Lambda)$ be a sequence which converges to zero vector
weakly in $L^{2}(\Lambda)$. We put $v_{n}=(H_{0,
\theta}+1)^{-1}u_{n}$. Since $(H_{0,\theta}+1)^{-1}$ is a bounded
operator from $L^{2}(\Lambda)$ to $H^{2}(\Lambda)$ and the
operator $H^{2}(\Lambda)\owns f\mapsto f|_{\Gamma ((0,L))}\in
L^{2}(\Gamma((0,L)))$ is compact, we have
$$ \Vert C_{\theta}^{1/2}(H_{0,\theta}+1)^{-1}u_{n}\Vert_
{L^{2}(\Lambda)}^{2}=c_{\theta}(v_{n},v_{n})= \Vert
v_{n}\Vert^{2}_{L^{2}(\Gamma((0,L)))}\rightarrow 0
\quad\mathrm{as}\quad n\rightarrow\infty.$$
Thus $C_{\theta}^{1/2}(H_{0,\theta}+1)^{-1}$ is a compact operator
 on $L^{2}(\Lambda)$, and consequently
$$ (H_{0,\theta}+1)^{-1}C_{\theta}(H_{0,\theta}+1)^{-1}
=[C_{\theta}^{1/2}(H_{0,\theta}+1)^{-1}]^{*}
[C_{\theta}^{1/2}(H_{0,\theta}+1)^{-1}]$$
is a compact operator on $L^{2}(\Lambda)$. \quad \QED \vspace{1em}
 \begin{lemma} \label{lemma3}
We have
$$ \sigma(H_{\beta})=\bigcup_{\theta\in[0,2\pi)}\sigma(
H_{\beta,\theta}).$$
\end{lemma}
{\sl Proof:} We put
$$ K_{\beta}=\int^{2\pi}_{0}\oplus H_{\beta,\theta}\,d\theta.$$
In view of Lemma 3.1, it suffices to prove that
 \begin{equation}
\sigma(K_{\beta})=\bigcup_{\theta\in[0,2\pi)}\sigma(
H_{\beta,\theta}).\label{3.11}
 \end{equation}
Combining Lemma~\ref{lemma2} with [RS, Theorem XIII.85(d)], we
 have
 \begin{equation}
\sigma(K_{\beta})\cap
[0,\infty)=\left(\bigcup_{\theta\in[0,2\pi)}\sigma(
H_{\beta,\theta})\right) \cap [0,\infty)=[0,\infty).\label{3.12}
 \end{equation}
Next we shall show that
 \begin{equation} \sigma(K_{\beta})\cap
(-\infty,0)=\left(\bigcup_{\theta\in[0,2\pi)}\sigma(
H_{\beta,\theta})\right) \cap (-\infty,0).\label{3.13}
 \end{equation}
For $n\in\mathbb{N}$, we put
$$ \alpha_{n}(\beta,\theta)= \sup_{v_{1},\cdots,v_{n-1}\in
L^{2}(\Lambda)}\; \inf_{\phi\in\mathcal{P}(v_1,\cdots,v_{n-1})}
q_{\beta,\theta}(\phi,\phi),$$
where $\mathcal{P}(v_1,\cdots,v_{n-1}):= \{ \phi;\, \phi\in
Q_{\theta},\, \Vert\phi\Vert_{L^{2}(\Lambda)}=1,\, \mathrm{and}\;
(\phi,v_{j})_{L^{2}(\Lambda)}=0 \;\;\mathrm{for}\;$ $1\leq j\leq
n-1 \}$. In order to prove $(3.13)$, we shall show that the
functions $\alpha_{n}(\beta,\cdot)$ are continuous on $[0,2\pi]$.
Let $\theta,\theta_{0}\in [0,2\pi]$. We define
$$ (V_{\theta,\theta_{0}}f)(x,y)=\exp\left\{{i\frac{\theta
-\theta_{0}}{K_{1}}x} \right\} f(x,y)\quad\mathrm{for}\quad f\in
L^{2}(\Lambda).$$
Then $V_{\theta,\theta_{0}}$ is a unitary operator on
$L^{2}(\Lambda)$ which maps $Q_{\theta_{0}}$ onto $Q_{\theta}$
bijectively. We have
\begin{eqnarray}
\lefteqn{q_{\beta,\theta}
(V_{\theta,\theta_{0}}g,V_{\theta,\theta_{0}}g)
-q_{\beta,\theta_{0}}(g,g)} \nonumber \\ &&
=\frac{(\theta-\theta_{0})^{2}}{K^{2}_{1}} \Vert
g\Vert^{2}_{L^{2}(\Lambda)} +2\Re
\left(i\frac{\theta-\theta_{0}}{K_{1}}V_{\theta,\theta_{0}}g,\,
e^{i\frac{\theta-\theta_{0}}{K_{1}}x} \frac{\partial}{\partial
x}g\right)_{L^{2}(\Lambda)}\label{3.14}
\end{eqnarray}
for $g\in Q_{\theta_{0}}$. Note that there exists $\alpha>0$ such
that
$$ \left\Vert\frac{\partial}{\partial x}g
\right\Vert^{2}_{L^{2}(\Lambda)}
\leq\frac{3}{2}q_{\beta,\theta_{0}}(g,g)+\alpha \Vert
g\Vert^{2}_{L^{2}(\Lambda)}\quad\mathrm{for}\quad g\in
Q_{\theta_{0}}.$$
Combining this with $(3.14)$, we obtain
 \begin{eqnarray*} \lefteqn{
|q_{\beta,\theta}(V_{\theta,\theta_{0}}g,V_{\theta,
\theta_{0}}g)
-q_{\beta,\theta_{0}}(g,g)|} \\ && \leq
\frac{(\theta-\theta_{0})^{2}}{K^{2}_{1}}\Vert g\Vert^{2}
_{L^{2}(\Lambda)}+\frac{|\theta-\theta_{0}|}{K_{1}} \left(
(1+\alpha)\Vert g\Vert^{2}_{L^{2}(\Lambda)}
+\frac{3}{2}q_{\beta,\theta_{0}}(g,g) \right)
\end{eqnarray*}
for $g\in Q_{\theta_{0}}$. It proves the continuity of
$\alpha_{n}(\beta,\cdot)$ on $[0,2\pi]$. Combining this with the
min-max principle and [RS, Theorem XIII.85(d)], we arrive at
(\ref{3.13}). The relations (\ref{3.12}) and (\ref{3.13}) together
give (\ref{3.11}) which completes the proof.\quad \QED
\vspace{1em}

The most important part of the proof is the analysis of the
discrete spectrum of $H_{\beta,\theta}$. The tool we use is the
Dirichlet-Neumann bracketing. Given $a>0$, we put
$$\Sigma_{a}=\Phi((0,L)\times (-a,a)).$$
Note that $\Sigma_{a}$ is a domain derived by transporting a
segment of the length $2a$ perpendicular to $\Gamma$ along the
curve. Since $\Gamma^{\prime}(0)=\Gamma^{\prime}(L)=(1,0)$, we
have $\Phi_{1}(0,\cdot)=0$ and $\Phi_{1}(L,\cdot)=K_1$ on
${\mathbb R}$. This together with $(A.5)$ implies, for
$|a|<a_{0}$, that $\Sigma_{a}\subset\Lambda$ and that
$\Lambda\backslash\Sigma_{a}$ consists of two connected
components, which we denote by $\Lambda^{1}_{a}$ and
$\Lambda^{2}_{a}$. For $\theta\in [0,2\pi)$, we define
 \begin{eqnarray*}
R^{+}_{a,\theta} &=& \{u\in H^{1}(\Sigma_{a});\quad
u=0\quad\mathrm{on}\quad\partial\Sigma_{a}\cap\Lambda,\\
&{}&\phantom{AAAAAAAAA}u(K_{1},\cdot) =
e^{i\theta}u(0,\cdot)\quad\mathrm{on}\quad (-a,a)\}, \\
R^{-}_{a,\theta} &=& \{ u\in H^{1}(\Sigma_{a});\quad
u(K_{1},\cdot)=e^{i\theta}u(0,\cdot)\quad\mathrm{on}\quad
(-a,a)\}, \\ q_{a,\beta,\theta}^{+}(f,f) &=& \Vert\nabla
f\Vert^{2}_{L^{2}(\Sigma_{a})}
-\beta\int_{\Gamma((0,L))}|f(x)|^{2} \,dS\quad \mathrm{for}\quad
f\in R^{+}_{a,\theta}, \\ q_{a,\beta,\theta}^{-}(f,f) &=&
\Vert\nabla f\Vert^{2}_{L^{2}(\Sigma_{a})}
-\beta\int_{\Gamma((0,L))}|f(x)|^{2} \,dS\quad\mathrm{for}\quad
f\in R^{-}_{a,\theta}.
 \end{eqnarray*}
Let $L_{a,\beta,\theta}^{+}$ and $L_{a,\beta,\theta}^{-}$ be the
self-adjoint operators associated with the forms
$q_{a,\beta,\theta}^{+}$ and $q_{a,\beta,\theta}^{-}$,
respectively. For $j=1,2$, we define
 \begin{eqnarray*} K^{+}_{a,j,\theta} &=& \{f\in
H^{1}(\Lambda^{j}_{a});\quad
f(K_{1},K_{2}+u)=e^{i\theta}f(0,u)\quad\mathrm{if}\quad
(0,u)\in\partial\Lambda^{j}_{a},\\ && \phantom{AAAAAAAAA}
f=0\quad\mathrm{on}\quad
\partial\Lambda^{j}_{a}\cap\Lambda\},
\\ K^{-}_{a,j,\theta} &=& \{f\in H^{1}(\Lambda^{j}_{a});\quad
f(K_{1},K_{2}+u)=e^{i\theta}f(0,u)\quad\mathrm{if}\quad
(0,u)\in\partial\Lambda^{j}_{a}\},
\\ && \phantom{AAAAAAAAA} e^{\pm}_{a,j,\theta}(f,f)
=\Vert\nabla f\Vert^{2}_
{L^{2}(\Lambda^{j}_{a})}\quad\mathrm{for}\quad f\in
K^{\pm}_{a,j,\theta}.
 \end{eqnarray*}
Let $E^{\pm}_{a,j,\theta}$ be the self-adjoint operators associated
with the forms $e^{\pm}_{a,j,\theta}$. By the bracketing bounds
(see [RS, XIII.15, Proposition 4]) we obtain
 \begin{equation}
E^{-}_{a,1,\theta} \oplus L_{a,\beta,\theta}^{-} \oplus
E^{-}_{a,2,\theta} \leq H_{\beta,\theta}\leq E^{+}_{a,1,\theta}
\oplus L_{a,\beta,\theta}^{+} \oplus E^{+}_{a,2,\theta}
\label{3.15}
 \end{equation}
in $\quad L^{2}(\Lambda^{1}_{a}) \oplus L^{2}(\Sigma_{a})\oplus
L^{2}(\Lambda^{2}_{a})$. In order to estimate the negative
eigenvalues of $H_{\beta,\theta}$, it is sufficient to estimate
those of $L_{a,\beta,\theta}^{+}$ and $L_{a,\beta,\theta}^{-}$
because the other operators involved in (\ref{3.15}) are
non-negative.

To this aim we introduce two operators in $L^{2}((0,L)\times
(-a,a))$ which are unitarily equivalent to
$L^{+}_{a,\beta,\theta}$ and $L^{-}_{a,\beta,\theta}$,
respectively. We define
 \begin{eqnarray*}
&& Q_{a,\theta}^{+} = \{\varphi\in H^{1}((0,L)\times (-a,a));\quad
\varphi(K_{1},\cdot)=e^{i\theta}\varphi(0,\cdot)
\quad\mathrm{on}\quad (-a,a), \\ && \phantom{AAAAAAAAAAAAAAA}
\varphi(\cdot ,a)=\varphi(\cdot, -a)=0 \quad\mathrm{on}\quad
(0,L)\}, \\ && Q_{a,\theta}^{-} = \{\varphi\in H^{1}((0,L)\times
(-a,a)); \quad \varphi(K_{1},\cdot)= e^{i\theta}
\varphi(0,\cdot)\quad\mathrm{on}\quad (-a,a)\}, \\ &&
b^{+}_{a,\beta,\theta}(f,f) = \int^{L}_{0}
\int^{a}_{-a}(1+u\gamma(s))^{-2} \left|\frac{\partial f}{\partial
s}\right|^{2}\,duds +\int^{L}_{0}\int^{a}_{-a}
\left|\frac{\partial f}{\partial u}\right|^{2}\,duds \\ &&
\phantom{AAAAAA} +\int^{L}_{0}\int^{a}_{-a}V(s,u)|f|^{2}\,dsdu
-\beta\int^{L}_{0}|f(s,0)|^{2}\,ds \quad\mathrm{for}\; f\in
Q^{+}_{a,\theta}, \\ && b^{-}_{a,\beta,\theta}(f,f) =
\int^{L}_{0}\int^{a}_{-a}(1+u\gamma(s))^{-2} \left|\frac{\partial
f}{\partial s}\right|^{2}\,duds +\int^{L}_{0}\int^{a}_{-a}
\left|\frac{\partial f}{\partial u}\right|^{2}\,duds \\ &&
\phantom{AAAAAA} +\int^{L}_{0}\int^{a}_{-a}V(s,u)|f|^{2}\,dsdu
-\beta\int^{L}_{0}|f(s,0)|^{2}\,ds \\ && \phantom{AAAAAA}
-\frac{1}{2}\int^{L}_{0}
\frac{\gamma(s)}{1+a\gamma(s)}|f(s,a)|^{2}\,ds
+\frac{1}{2}\int^{L}_{0}
\frac{\gamma(s)}{1-a\gamma(s)}|f(s,-a)|^{2}\,ds
\end{eqnarray*}
for $f\in Q^{-}_{a,\theta}$, where
$$ V(s,u)= \frac{1}{2}(1+u\gamma(s))^{-3}u\gamma^{\prime\prime}(s)
-\frac{5}{4}(1+u\gamma(s))^{-4}u^{2}\gamma^{\prime}(s)^{2}
-\frac{1}{4}(1+u\gamma(s))^{-2}\gamma(s)^{2}.$$
Let $B^{+}_{a,\beta,\theta}$ and $B^{-}_{a,\beta,\theta}$ be the
self-adjoint operators associated with the forms
$b^{+}_{a,\beta,\theta}$ and $b^{-}_{a,\beta,\theta}$,
respectively. Acting as in the proof of Lemma 2.2 in \cite{EY}, we
arrive at the following result.
 \begin{lemma} \label{lemma4}
The operators $B^{+}_{a,\beta,\theta}$ and
$B^{-}_{a,\beta,\theta}$ are unitarily equivalent to
$L^{+}_{a,\beta,\theta}$ and $L^{-}_{a,\beta,\theta}$,
respectively.
\end{lemma}
Next we estimate $B^{+}_{a,\beta,\theta}$ and
$B^{-}_{a,\beta,\theta}$ by operators with separated variables. We
put
$$\gamma_{+}=\max_{[0,L]}|\gamma(\cdot)|,\quad
\gamma^{\prime}_{+} =\max_{[0,L]}|\gamma^{\prime}(\cdot)|,\quad
\gamma^{\prime\prime}_{+}=\max_{[0,L]}|
\gamma^{\prime\prime}(\cdot)|,$$
and
\begin{eqnarray*} V_{+}(s) &=& \frac{1}{2}(1-a\gamma_{+})^{-3}
a\gamma^{\prime\prime}_{+} -\frac{5}{4}(1+a\gamma_{+})^{-4}
a^{2}(\gamma^{\prime}_{+})^{2}
-\frac{1}{4}(1+a\gamma_{+})^{-2}\gamma(s)^{2}, \\ V_{-}(s) &=&
-\frac{1}{2}(1-a\gamma_{+})^{-3} a\gamma^{\prime\prime}_{+}
-\frac{5}{4}(1-a\gamma_{+})^{-4} a^{2}(\gamma^{\prime}_{+})^{2}
-\frac{1}{4}(1-a\gamma_{+})^{-2}\gamma(s)^{2}. \end{eqnarray*}
If $0<a<\frac{1}{2\gamma_{+}}$, we can define
\begin{eqnarray*}
\tilde{b}^{+}_{a,\beta,\theta}(f,f)&=&
(1-a\gamma_{+})^{-2}\int^{L}_{0}\int^{a}_{-a} \left|\frac{\partial
f}{\partial s}\right|^{2}\,duds +\int^{L}_{0}\int^{a}_{-a}
\left|\frac{\partial f}{\partial u}\right|^{2}\,duds\\
&&+\int^{L}_{0}\int^{a}_{-a}V_{+}(s)|f|^{2}\,duds
-\beta\int^{L}_{0}|f(s,0)|^{2}\,ds \quad\mathrm{for}\;\; f\in
Q^{+}_{a,\theta}, \\ \tilde{b}^{-}_{a,\beta,\theta}(f,f)&=&
(1+a\gamma_{+})^{-2}\int^{L}_{0}\int^{a}_{-a} \left|\frac{\partial
f}{\partial s}\right|^{2}\,duds +\int^{L}_{0}\int^{a}_{-a}
\left|\frac{\partial f}{\partial u}\right|^{2}\,duds\\
&&+\int^{L}_{0}\int^{a}_{-a}V_{-}(s)|f|^{2}\,duds
-\beta\int^{L}_{0}|f(s,0)|^{2}\,ds\\
&&-\gamma_{+}\int^{L}_{0}(|f(s,a)|^{2}+|f(s,-a)|^{2})\,ds
\quad\mathrm{for}\quad f\in Q^{-}_{a,\theta}.
\end{eqnarray*}
Then we have
  \begin{equation} b^{+}_{a,\beta,\theta}(f,f) \leq
\tilde{b}^{+}_{a,\beta,\theta}(f,f) \quad\mathrm{for}\quad f\in
Q^{+}_{a,\theta},\label{3.16} \end{equation}
\begin{equation}
\tilde{b}^{-}_{a,\beta,\theta}(f,f)\leq
 {b}^{-}_{a,\beta,\theta}(f,f)
\quad\mathrm{for}\quad f\in Q^{-}_{a,\theta}.\label{3.17}
\end{equation}
Let $\tilde{H}^{+}_{a,\beta,\theta}$ and
$\tilde{H}^{-}_{a,\beta,\theta}$ be the self-adjoint operators
associated with the forms $\tilde{b}^{+}_{a,\beta,\theta}$ and
$\tilde{b}^{-}_{a,\beta,\theta}$, respectively. Let
$T^{+}_{a,\beta}$ be the self-adjoint operator associated with the
form
$$t^{+}_{a,\beta}(f,f)= \int^{a}_{-a}|f^{\prime}(u)|^{2}\,du-\beta
|f(0)|^{2},\quad f\in H^{1}_{0}((-a,a)).$$
Let finally $T^{-}_{a,\beta}$ be the self-adjoint operator
associated with the form
$$t^{-}_{a,\beta}(f,f) =\int^{a}_{-a}|f^{\prime}(u)|^{2}\,du-\beta
|f(0)|^{2} -\gamma_{+}(|f(a)|^{2}+|f(-a)|^{2})$$
for $f\in H^{1}((-a,a))$. We define
\begin{eqnarray*}
U^{+}_{a,\theta} &=& -(1-a\gamma_{+})^{-2} \frac{d^{2}}{d
s^{2}}+V_{+}(s)\quad\mathrm{in}\quad L^{2}((0,L))
\quad\mathrm{with}\,\,\mathrm{the}\,\,\mathrm{domain}\quad
P_{\theta}, \\ U^{-}_{a,\theta} &=& -(1+a\gamma_{+})^{-2}
\frac{d^{2}}{d s^{2}}+V_{-}(s) \quad\mathrm{in}\quad L^{2}((0,L))
\quad\mathrm{with}\,\,\mathrm{the}\,\,\mathrm{domain}\quad
P_{\theta}. \end{eqnarray*}
Then we have
  \begin{eqnarray}
  \tilde{H}^{+}_{a,\beta,\theta}
=U^{+}_{a,\theta}\otimes 1+1\otimes T^{+}_{a,\beta}, \nonumber \\
\tilde{H}^{-}_{a,\beta,\theta} =U^{-}_{a,\theta}\otimes 1+1\otimes
T^{-}_{a,\beta}.\label{(3.18)} \end{eqnarray}
Next we consider the asymptotic behaviour for a fixed eigenvalue
of $U^{\pm}_{a,\theta}$ as $a$ tends to zero. Let
$\mu^{\pm}_{j}(a,\theta)$ be the $j$th eigenvalue of
$U^{\pm}_{a,\theta}$ counted with multiplicity. We recall the
estimates contained in relations $(2.25)$ and $(2.26)$ of the
paper \cite{Yo}.
 \begin{proposition} \label{prop5}
For $j\in \mathbb{N}$ and $0<a<\frac{1}{2\gamma_{+}}$, there
exists $C_{j}>0$ such that
$$|\mu^{+}_{j}(a,\theta) -\mu_{j}(\theta)|\leq C_{j}a
\label{3.19}$$
and
$$|\mu^{-}_{j}(a,\theta) -\mu_{j}(\theta)|\leq C_{j}a,
\label{3.20}$$
where $C_{j}$ is independent of $a$ and $\theta$.
\end{proposition}
We also need two-sided estimates for the first eigenvalue of the
transverse operators $T^{\pm}_{a,\beta}$. They are obtained in the
same way as in \cite{EY}: we get
 \begin{proposition} \label{prop6}
Assume that $\beta a>\frac{8}{3}$. Then $T^{+}_{a,\beta}$ has only
one negative eigenvalue, which we denote by $\zeta^{+}_{a,\beta}$.
It satisfies the inequalities
$$-\frac{1}{4}\beta^{2}<\zeta^{+}_{a,\beta}<
-\frac{1}{4}\beta^{2}+2\beta^{2}\exp \left(-\frac{1}{2}\beta
a\right).$$
\end{proposition}
 \begin{proposition} \label{prop7}
Let $\beta a>8$ and $\beta>\frac{8}{3}\gamma_{+}$. Then
$T^{-}_{a,\beta}$ has a unique negative eigenvalue
$\zeta^{-}_{a,\beta}$, and moreover, we have
$$-\frac{1}{4}\beta^{2}-
\frac{2205}{16}\beta^{2}\exp\left(-\frac{1}{2}\beta a\right)
<\zeta^{-}_{a,\beta}<-\frac{1}{4}\beta^{2}.$$
\end{proposition}

Now we are ready to prove our main result. \\[1em]
 {\sl Proof of Theorem~\ref{main}:} We put $a(\beta)
=6\beta^{-1}\log\beta$. Let $\xi^{\pm}_{\beta,j}$ be the $j$th
eigenvalue of $T^{\pm}_{a(\beta),\beta}$. From Propositions 3.6
and 3.7, we have
$$\xi^{\pm}_{\beta,1}
=\zeta^{\pm}_{a(\beta),\beta}\quad\mathrm{and}\quad
\xi^{\pm}_{\beta,2}\geq 0.$$
From $(3.18)$, we infer that $\{\xi^{\pm}_{\beta,j}
+\mu^{\pm}_{k}(a(\beta),\theta)\}_{j,k\in\mathbb{N}}$, properly
ordered, is the sequence of all eigenvalues of
$\tilde{H}^{\pm}_{a(\beta),\beta,\theta}$ counted with
multiplicity. Using Proposition~3.5, we find
$$\xi^{\pm}_{\beta,j}
+\mu^{\pm}_{k}(a(\beta),\theta)\geq\mu^{\pm}_{1}(a(\beta),\theta)
=\mu_{1}(\theta)+\mathcal{O}(\beta^{-1}\log\beta)\eqno{(3.21)}$$
for $j\geq 2$ and $k\geq 1$, where the error term is uniform with
respect to the quasimomentum $\theta\in [0,2\pi)$. For
$k\in\mathbb{N}$ and $\theta\in [0,2\pi)$, we define
$$\tau^{\pm}_{\beta,k,\theta} =\zeta^{\pm}_{a(\beta),\beta}
+\mu^{\pm}_{k}(a(\beta),\theta).\eqno{(3.22)}$$
From Propositions 3.5--3.7 we get
$$\tau^{\pm}_{\beta,k,\theta}= -\frac{1}{4}\beta^{2}
+\mu_{k}(\theta) +\mathcal{O}(\beta^{-1}\log\beta)
\quad\mathrm{as} \quad\beta\rightarrow\infty,\eqno{(3.23)}$$
where the error term is uniform with respect to $\theta\in
[0,2\pi)$. Let $n\in\mathbb{N}$. Combining $(3.21)$ with $(3.23)$,
we claim that there exists $\beta(n)>0$ such that
$$ \tau^{+}_{\beta,n,\theta}<0,
\quad\tau^{+}_{\beta,n,\theta}<\xi^{+}_{\beta,j}
+\mu^{+}_{k}(a(\beta),\theta),\quad\mathrm{and}\quad
\tau^{-}_{\beta,n,\theta}<\xi^{-}_{\beta,j}
+\mu^{-}_{k}(a(\beta),\theta)$$
for $\beta\geq\beta(n)$, $j\geq 2$, $k\geq 1$, and $\theta\in
[0,2\pi)$. Hence the $j$th eigenvalue of
$\tilde{H}^{\pm}_{a(\beta),\beta,\theta}$ counted with
multiplicity is $\tau^{\pm}_{\beta,j,\theta}$ for $j\leq n$,
$\beta\geq\beta(n)$, and $\theta\in [0,2\pi)$. Let $\beta\geq
\beta(n)$ and denote by $\kappa^{\pm}_{j}(\beta,\theta)$ the $j$th
eigenvalue of $L^{\pm}_{a(\beta),\beta,\theta}$. From $(3.16)$,
$(3.17)$, and the min-max principle, we obtain
$$\tau^{-}_{\beta,j,\theta}\leq\kappa^{-}_{j}(\beta,\theta)
\quad\mathrm{and}\quad\kappa^{+}_{j}(\beta,\theta)\leq
\tau^{+}_{\beta,j,\theta}\quad\mathrm{for}\quad 1\leq j\leq n,
\eqno{(3.24)}$$
so we have $\kappa^{+}_{n}(\beta,\theta)<0$. Hence the min-max
principle and $(3.15)$ imply that $H_{\beta,\theta}$ has at least
$n$ eigenvalues in $(-\infty,\kappa^{+}_{n}(\beta,\theta))$. For
$1\leq j\leq n$, we denote by $\lambda_{j}(\beta,\theta)$ the
$j$th eigenvalue of $H_{\beta,\theta}$. We have
$$\kappa^{-}_{j}(\beta,\theta)\leq
\lambda_{j}(\beta,\theta)\leq\kappa^{+}_{j}(\beta,\theta)
\quad\mathrm{for}\quad 1\leq j\leq n.$$
This together with $(3.23)$ and $(3.24)$ implies that
$$\lambda_{j}(\beta,\theta)=-\frac{1}{4}\beta^{2}
+\mu_{j}(\theta)+\mathcal{O}(\beta^{-1}\log\beta)
\quad\mathrm{as}\quad \beta\rightarrow\infty\quad\mathrm{for}\quad
1\leq j\leq n,$$
where the error term is uniform with respect to $\theta\in
[0,2\pi)$, and completes thus the proof of Theorem~\ref{main}.
\QED \vspace{1em}

Our next aim is to prove Corollary~2.2. As a preliminary, we
denote by $B_{j}$ and $G_{j}$, respectively, the length of the
$j$th band and the $j$th gap
 of the spectrum of the operator $-\frac{d^{2}}{ds^{2}}
-\frac{1}{4}\gamma(s)^{2}$ in $L^{2}(\mathbb{R})$ with the domain
$H^{2}(\mathbb{R})$:
\begin{eqnarray*} B_{j} &=& \cases{
\mu_{j}(\pi)-\mu_{j}(0)&\rm{for}\,\,\rm{odd}\quad $j$,\cr
\mu_{j}(0)-\mu_{j}(\pi)&\rm{for}\,\,\rm{even}\quad $j$,\cr}
\\ G_{j} &=& \cases{\mu_{j+1}(\pi)-\mu_{j}(\pi)&\rm{for}
\,\,\rm{odd}\quad
$j$,\cr \mu_{j+1}(0)-\mu_{j}(0)&\rm{for}\,\,\rm{even}\quad $j$.\cr}
\end{eqnarray*}
Since $\mu_{j}(\cdot)$ is continuous on $[0,2\pi]$, we immediately
obtain from Theorem~\ref{main} the following claim.
 \begin{lemma} \label{lemma18}
For $n\in\mathbb{N}$, we have
\begin{eqnarray*} |\lambda_{n}(\beta,[0,2\pi))|=B_{n}+\mathcal{O}(
\beta^{-1}\log\beta) &\quad\mathrm{as}\quad&
\beta\rightarrow\infty,
\\ \min_{\theta\in [0,2\pi)}\lambda_{n+1}(\beta,\theta)
-\max_{\theta\in [0,2\pi)}\lambda_{n}(\beta,\theta)
=G_{n}+\mathcal{O}( \beta^{-1}\log\beta) &\quad\mathrm{as}\quad&
\beta\rightarrow\infty.
\end{eqnarray*}
\end{lemma}
Now we recall Borg's theorem (see [Bo, Ho, Un]).
 \begin{theorem} \label{Borg} {\rm (Borg)}
Suppose that $W$ is a real-valued, piecewise continuous function
on $[0,L]$. Let $\alpha^{\pm}_{j}$ be the $j$th eigenvalue of the
following operator counted with multiplicity:
$$-\frac{d^{2}}{ds^{2}}+W(s)\quad\mathrm{in}\quad L^{2}((0,L))$$
with the domain
$$\{v\in H^{2}((0,L));\quad v(L)=\pm v(0),\quad v^{\prime}(L)=\pm
v^{\prime}(0)\}.$$
Suppose that
$$\alpha^{+}_{j}=\alpha^{+}_{j+1}\quad\mathrm{for}\,\,\mathrm{all}
\,\,\mathrm{even}\quad j,$$
and
$$\alpha^{-}_{j}=\alpha^{-}_{j+1}
\quad\mathrm{for}\,\,\mathrm{all}\,\,
\mathrm{odd}\quad j.$$
Then $W$ is constant on $[0,L]$.
\end{theorem}

\noindent{\sl Proof of Corollary~2.2:} Assume that $\gamma$ is not
identically zero. Then it follows from $(A.3)$ that $\gamma$ is
not constant on $[0,L]$. Combining this with Borg's theorem, we
infer that there exists $m\in\mathbb{N}$ such that $G_{m}>0$. From
Lemma~3.8 we get
$$\lim_{\beta\rightarrow\infty}\left( \min_{\theta\in
[0,2\pi)}\lambda_{m+1}(\beta,\theta) -\max_{\theta\in
[0,2\pi)}\lambda_{m}(\beta,\theta)\right) =G_{m}>0.$$
This completes the proof. \QED


\setcounter{equation}{0}
\section{The gaps of Hill's equation}

It follows from Lemma 3.8 that if the $m$th gap of
$-\frac{d^{2}}{ds^{2}} -\frac{1}{4}\gamma(s)^{2}$ in
$L^{2}(\mathbb{R})$ is open, so is the $m$th gap of $H(\beta)$ for
sufficiently large $\beta>0$. It is thus useful to find a
sufficient condition for which the $m$th gap of our comparison
operator is open for a given $m\in \mathbb{N}$. Since a particular
form of the effective potential is not essential, we will do that
for gaps of the Hill operator with a general bounded potential.

Let $V\in L^{\infty}((-a/2,a/2))$ and denote by
$\{a_{j}\}^{\infty}_{j=1}$ and $\{b_{j}\}^{\infty}_{j=0}$ the
sequences of its Fourier coefficients:
$$V(x)=\sum^{\infty}_{j=1}a_{j}\sin\frac{2\pi j}{a}x
+\sum^{\infty}_{j=0}b_{j}\cos\frac{2\pi j}{a}x\quad\mathrm{in}
\quad L^{2}((-a/2,a/2)),$$
where
 \begin{eqnarray*}
 a_{j} &=& \frac{2}{a}\int^{a/2}_{-a/2}V(x) \sin\frac{2\pi
j}{a}x\,dx, \\ b_{j} &=& \frac{2}{a}\int^{a/2}_{-a/2}V(x)
\cos\frac{2\pi j}{a}x\,dx. \end{eqnarray*}
Let $\kappa_{j}$ be the $j$th eigenvalue of the operator
$$-\frac{d^{2}}{ds^{2}}+V(x)\quad\mathrm{in}\quad
L^{2}((-a/2,a/2))\quad\mathrm{with} \,\,\mathrm{periodic}\,\,
\mathrm{b.c.}, \eqno{(4.1)}$$
and similarly, let $\nu_{j}$ be the $j$th eigenvalue of the
operator
$$-\frac{d^{2}}{ds^{2}}+V(x)\quad\mathrm{in}\quad
L^{2}((-a/2,a/2))\quad\mathrm{with}\,\,\mathrm{antiperiodic}\,\,
\mathrm{b.c.}. \eqno{(4.2)}$$

We are going to prove the following result.
 \begin{theorem} \label{thm1}
Let $n\in\mathbb{N}$. Assume that
$$0<\sqrt{a^{2}_{n}+b^{2}_{n}}<\frac{12\pi^{2}}{a^{2}}n^{2}$$
and
$$\left\Vert V(x)-b_{0}-a_{n}\sin\frac{2\pi  n}{a}x
-b_{n}\cos\frac{2\pi n}{a}x\right\Vert_ {L^{\infty}((-a/2,a/2))}<
\frac{1}{4}\sqrt{a^{2}_{n}+b^{2}_{n}}.$$
Then we have
$$\nu_{n+1}-\nu_{n}>0 \quad\mathrm{when}\quad
n\quad\mathrm{is}\,\,\mathrm{odd},$$
 and
 $$\kappa_{n+1}-\kappa_{n}>0\quad\mathrm{when}\quad
 n\quad\mathrm{is}\,\,\mathrm{even}.$$
\end{theorem}
\noindent Proposition 2.3 immediately follows from Theorem
 4.1. In order to prove the latter, we shall estimate the
 length of the first gap of the Mathieu operator. For $\alpha
\in\mathbb{R}$, we define
$$M_{\alpha}=-\frac{d^{2}}{dx^{2}}+2\alpha
\cos\frac{2\pi}{a}x\quad\mathrm{in}\quad L^{2}((-a/2,a/2))$$
with the domain
$$D=\{u\in H^{2}((-a/2,a/2));\quad u(a/2)=-u(-a/2),\;
u^{\prime}(a/2)=-u^{\prime}(-a/2)\}.$$
By $m_{j}(\alpha)$ we denote the $j$th eigenvalue of $M_{\alpha}$
counted with multiplicity. The sought estimate looks as follows:
 \begin{theorem} \label{thm2}
We have
$$m_{2}(\alpha)-m_{1}(\alpha)\geq |\alpha|
\quad\mathrm{provided}\,\,\mathrm{that}\quad |\alpha|
<\frac{6\pi^{2}}{a^{2}}.$$
\end{theorem}
{\sl Proof:} We prove the assertion only for $\alpha<0$ because
that for $\alpha>0$ is similar. We put
\begin{eqnarray*} D^{+} &=& \{u\in H^{2}((0,a/2)); \quad
u^{\prime}(0)=u(a/2)=0\}, \\ D^{-} &=& \{u\in H^{2}((0,a/2));
\quad u(0)=u^{\prime}(a/2)=0\}\end{eqnarray*}
and define
$$L^{\pm}_{\alpha}=-\frac{d^{2}}{dx^{2}}+2\alpha\cos\frac{2\pi}{a}x
\quad\mathrm{in}\quad L^{2}((0,a/2))\quad\mathrm{with}\,\,
\mathrm{the}\,\,\mathrm{domain}\quad D^{\pm}.$$
By $\mu^{\pm}_{1}(\alpha)$ we denote the first eigenvalue of
$L^{\pm}_{\alpha}$. Since the function $\cos\frac{2\pi}{a}x$ is
even, we infer that $M_{\alpha}$ is unitarily equivalent to the
operator $L^{+}_{\alpha}\oplus L^{-}_{\alpha}$ in $L^{2}((0,a/2))
\oplus L^{2}((0,a/2))$. We put
$$\varphi_{j}(x)=\frac{2}{\sqrt{a}}\sin\frac{\pi}{a}(2j-1)x
\quad\mathrm{and}\quad
\psi_{j}(x)=\frac{2}{\sqrt{a}}\cos\frac{\pi}{a}(2j-1)x.$$
It is clear that
$$\{\varphi_{j}\}^{\infty}_{j=1}\subset D^{-}
\quad\mathrm{and}\quad \{\psi_{j}\}^{\infty}_{j=1}\subset D^{+},$$
and, in addition, $\{\varphi_{j}\}^{\infty}_{j=1}$ and
$\{\psi_{j}\}^{\infty}_{j=1}$ are complete orthonormal systems of
$L^{2}((0,a/2))$. We first estimate $\mu^{+}_{1}(\alpha)$ from
above. By the min-max principle, we obtain
$$\mu^{+}_{1}(\alpha)\leq (L^{+}_{\alpha}\psi_{1},\psi_{1})
=\left(\frac{\pi}{a}\right)^{2}+\alpha.\eqno{(4.3)}$$
Next we estimate $\mu^{-}_{1}(\alpha)$ from below. Let $\phi\in
D^{-}$ and $\Vert\phi\Vert_{L^{2}((0,a/2))}=1$. Since
$\{\varphi_{j}\}^{\infty}_{j=1}$ is a complete orthonormal system
of $L^{2}((0,a/2))$, we have
$$\phi(x)=\sum^{\infty}_{j=1}s_{j}\varphi_{j},\quad
\sum^{\infty}_{j=1}s^{2}_{j}=1,$$
where $s_{j}=(\phi,\varphi_{j})_{L^{2}((0,a/2))}$ are the Fourier
coefficients. We have
\begin{eqnarray*}
\lefteqn{(L^{-}_{\alpha}\phi,\phi)_{L^{2}((0,a/2))}
-\left(\frac{\pi}{a}\right)^{2}\Vert\phi\Vert^{2}_{L^{2}((0,a/2))}
} \\
&&=\sum^{\infty}_{j=2}s^{2}_{j}\left(\frac{\pi}{a}\right)^{2}4j(j-1)
+\alpha \left(2\sum^{\infty}_{j=1}s_{j}s_{j+1}-s^{2}_{1}\right)\\
&&=\sum^{\infty}_{j=2}s^{2}_{j}\left(\frac{\pi}{a}\right)^{2}4j(j-1)
+\alpha \left[2\sum^{\infty}_{j=2}s_{j}s_{j+1}-
(s_{1}-s_{2})^{2}+s^{2}_{2}\right]\\
&&\geq\sum^{\infty}_{j=2}
s^{2}_{j}\left(\frac{\pi}{a}\right)^{2}4j(j-1)
+\alpha \left[2\sum^{\infty}_{j=2}s_{j}s_{j+1} +s^{2}_{2}\right]\\
&&\geq\sum^{\infty}_{j=2}s^{2}_{j}
\left(\frac{\pi}{a}\right)^{2}4j(j-1)
+\alpha \left[\sum^{\infty}_{j=2}
\left(\frac{1}{3}s_{j}^{2}+3s_{j+1}^{2}\right) +s^{2}_{2}\right]\\
&&=\left[ 8\left(\frac{\pi}{a}\right)^{2}+\frac{4}{3}\alpha\right]
s^{2}_{2}+ \sum^{\infty}_{j=3}\left[
\left(\frac{\pi}{a}\right)^{2}4j(j-1)+\frac{10}{3}\alpha\right]
s^{2}_{j}\\ &&\geq 0\quad\mathrm{for}\quad
-\frac{6\pi^{2}}{a^{2}}<\alpha<0.
\end{eqnarray*}
This together with the min-max principle implies that
$$\mu^{-}_{1}(\alpha)\geq \left(\frac{\pi}{a}\right)^{2}
\quad\mathrm{for}\quad
-\frac{6\pi^{2}}{a^{2}}<\alpha<0.\eqno{(4.4)}$$
Combining $(4.4)$ with $(4.3)$, we obtain the assertion of the
theorem. \QED \vspace{1em}

Now we are ready to prove the main result of this section. \\[1em]
{\sl Proof of Theorem 4.1:} We prove the assertion for odd $n$
only since the argument for even $n$ is similar. We extend $V$ to
an $a$-periodic function which we denote by $\tilde{V}$. Let
$\tau\in [0,2\pi)$ be such that
$$\cos\tau=\frac{b_{n}}{\sqrt{a_{n}^{2}+b_{n}^{2}}}
\quad\mathrm{and}\quad
\sin\tau=-\frac{a_{n}}{\sqrt{a_{n}^{2}+b_{n}^{2}}}.$$
We have
$$a_{n}\sin\frac{2\pi nx}{a}+b_{n}\cos\frac{2\pi nx}{a}
=\sqrt{a_{n}^{2}+b_{n}^{2}}\cos \frac{2n\pi}{a}\left(
x+\frac{a}{2n\pi}\tau\right).$$
Let $d_{j}$ be the $j$th eigenvalue of the operator with this
potential,
$$-\frac{d^{2}}{dx^{2}}+\sqrt{a_{n}^{2}+b_{n}^{2}}\cos
\frac{2n\pi}{a}\left(
x+\frac{a}{2n\pi}\tau\right)\quad\mathrm{in}\;\, L^{2}
\left(\left(-\frac{a}{2}-\frac{a}{2n\pi}\tau,
\frac{a}{2}-\frac{a}{2n\pi}\tau\right)\right)$$
with antiperiodic boundary condition. Since a coordinate shift
amounts to a unitary transformation and does not change the
spectrum, $d_{n+1}-d_{n}$ is equal to the difference of the first
two eigenvalues of the operator
 $$-\frac{d^{2}}{dx^{2}}+\sqrt{a_{n}^{2}+b_{n}^{2}}\cos\frac{2n\pi
 x}{a}\quad\mathrm{in}\quad
 L^{2}\left(\left(-\frac{a}{2n},\frac{a}{2n}\right)\right)$$
with antiperiodic boundary condition. Thus it follows from
Theorem~4.2 that
$$d_{n+1}-d_{n}\geq\frac{1}{2}\sqrt{a_{n}^{2}+b_{n}^{2}}.
\eqno{(4.5)}$$
Let $e_{j}$ be the $j$th eigenvalue of the operator
$$-\frac{d^{2}}{dx^{2}}+\tilde{V}(x)\quad\mathrm{in}\quad L^{2}
\left(\left(-\frac{a}{2}-\frac{a}{2n\pi}\tau,
\frac{a}{2}-\frac{a}{2n\pi}\tau \right)\right)$$
with antiperiodic boundary condition. By the min-max principle, we
get
$$|d_{j}-e_{j}|\leq
\left\Vert\tilde{V}(x)-b_{0}-\sqrt{a_{n}^{2}+b_{n}^{2}}
\cos\frac{2n\pi}{a} \left(x+\frac{a}{2n\pi}\tau\right)\right\Vert_
{L^{\infty}((-\frac{a}{2}-\frac{a}{2\pi}\tau,
\frac{a}{2}-\frac{a}{2\pi}\tau))}.\eqno{(4.6)}$$
Notice that $\nu_{j}=e_{j}$ for all $j\in \mathbb{N}$. This
together with $(4.5)$ and $(4.6)$ implies that
$\nu_{n+1}-\nu_{n}>0$, and completes therefore the proof of
Theorem~4.1. \QED


\setcounter{equation}{0}
\section{Asymptotically straight curves}

Finally, we are going to discuss briefly the case when $\Gamma$ is
non-periodic and asymptotically straight. We impose the following
assumptions on $\gamma$:
 \begin{description}
 \vspace{-.6ex}
\item{(A.6)} $\:\gamma\in C^{2}(\mathbb{R})$.
 \vspace{-1.2ex}
\item{(A.7)} The function $\gamma$ is not identically zero.
 \vspace{-1.2ex}
\item{(A.8)} There exists $c\in (0,1)$ such that $|\Gamma(s)-\Gamma
(t)|\geq c|t-s|$ for $s,t\in \mathbb{R}$.
 \vspace{-1.2ex}
\item{(A.9)} There exist $\tau>\frac{5}{4}$ and $K>0$ such that
$|\gamma(s)|\leq K|s|^{-\tau}$ for $s\in \mathbb{R}$.
 \vspace{-.6ex}
\end{description}
From [EI, Proposition 5.1 and Theorem 5.2] we know that under
these conditions
$$\sigma_{\mathrm{ess}}(H_{\beta})
=[-\frac{1}{4}\beta^{2},\infty)\quad\mathrm{and}\quad
\sigma_\mathrm{d}(H_{\beta})\neq\emptyset.$$
We define
$$S=-\frac{d^{2}}{ds^{2}}-\frac{1}{4}\gamma (s)^{2}
\quad\mathrm{in}\quad L^{2}(\mathbb{R}) \quad\mathrm{with}\,\,
\mathrm{the}\,\,\mathrm{domain}\quad H^{2}(\mathbb{R}).$$
Since $\gamma$ is not identically zero on $\mathbb{R}$, we have
$$\sigma_\mathrm{d}(S)\neq\emptyset$$
(see, e.g., [BGS] and [Si]). We put $n=\sharp\sigma_\mathrm{d}(S)$.
For $1\leq j\leq n$, we denote by $\mu_{j}$ the $j$th eigenvalue
of $S$ counted with multiplicity.
 \begin{theorem} \label{straight}
There exists $\beta_{0}>0$ such that
$\sharp\sigma_\mathrm{d}(H_{\beta})=n$ for $\beta \geq\beta_{0}.$
For $\beta\geq\beta_{0}$ and $1\leq j\leq n$, we denote by
$\lambda_{j}(\beta)$ the $j$th eigenvalue of $H_{\beta}$ counted
with multiplicity. Then we have
$$\lambda_{j}(\beta)=-\frac{1}{4}\beta^{2}+\mu_{j} +\mathcal
{O}(\beta^{-1}\log\beta)\quad\mathrm{as}\quad\beta\rightarrow
\infty\quad\mathrm{for}\quad 1\leq j\leq n.$$
\end{theorem}
\vspace{1em}
We omit the proof, since it analogous to those of Theorem 2.1 and
[EY, Theorem 1].

\subsection*{Acknowledments}

K.Y. appreciates the hospitality in NPI, Academy of Sciences, in
\v{R}e\v{z} where a part of this work was done. The research has
been partially supported by GAAS and the Czech Ministry of
Education within the projects A1048101 and ME170.

\end{document}